\documentclass[10pt]{iopart}

\usepackage{graphicx}

\begin{document}

\title[Epitaxial growth of C$_{60}$ on HOPG]{Epitaxial growth of C$_{60}$ on highly oriented pyrolytic graphite surfaces studied at low temperatures}

\author{E. Seydel$^1$, R. Hoffmann-Vogel$^{1,2}$, and M. Marz$^1$}
\address{$^1$ Physikalisches Institut, Karlsruhe Institute for Technology (KIT), 76131 Karlsruhe, Germany} 
\address{$^2$ Department of Physics, University of Konstanz, Universit\"{a}tstra\ss{}e 10, 78464 Konstanz, Germany}
\ead{Michael.Marz@kit.edu}
\ead{Regina.Hoffmann-Vogel@uni-konstanz.de}
\date{\today}

\begin{abstract}
Graphite surfaces interact weakly with molecules compared to other conducting surfaces bringing the molecule-molecule interaction to the foreground. C$_{60}$ on highly oriented pyrolytic graphite is a model system for studying the molecular self-assembly on surfaces. Our scanning tunneling microscopy measurements at liquid nitrogen temperatures confirm the previously observed island growth mode. Our results indicate that there is an epitaxial relationship of the molecular islands and the substrate with three possible orientations of the islands. For one of these orientations, we determine this epitaxial relationship by analyzing in detail an image taken across a C$_{60}$ island step edge. In this image we have obtained high resolution on both the molecular island and the substrate. The result of this analysis is confirmed by two-dimensional Fourier analysis.
\end{abstract}

\maketitle

\section{\label{sec:intro}Introduction}
Graphitic structures such as fullerenes are important model systems for the basic understanding of molecular self-assembly on surfaces. 
In many studies, substrates interacting weakly with the molecules are chosen to focus on the molecule-molecule interaction, for reviews see~\cite{Kuehnle2009,Hoffmann-Vogel2018}. 
On insulating surfaces this interaction has been investigated by using scanning force microscopy methods~\cite{Okita2001,Burke2005,Koerner2011}. On CaF$_2$ depending on the substrate temperature different growth modes have been observed. For high substrate temperatures larger than $310$~K triangular islands are formed, while for slightly lower temperatures below $300$~K hexagonal islands are found~\cite{Koerner2011}. On KBr dewetting with an unusual growth mode and shape of the C$_{60}$ islands has been observed~\cite{Burke2005,Burke2007}, whereas a wetting layer forms for C$_{60}$ on CaCO$_3$~\cite{Rahe2012}. These examples show the wide variety of growth modes of C$_{60}$ on insulating surfaces.

On metal surfaces, even for relatively inert substrates such as gold, C$_{60}$ molecules show a relatively strong interaction with the surface~\cite{Altman1992,Altman1993,Gaisch1994,Veenstra2002,Mativetsky2004}. Indeed an epitaxial relationship has been found between the C$_{60}$ islands and the Au(111) surface~\cite{Altman1992,Altman1993,Mativetsky2004}.
Highly oriented pyrolytic graphite (HOPG) is an important substrate to study the transition between conductive (strongly interacting) and insulating (weakly interacting) surfaces~\cite{Rahe2012}. C$_{60}$ grows on HOPG mainly at step edges in the form of round and elongated islands with hexagonally arranged edges~\cite{Szuba1999,Kenny2000,Liu2006,Liu2008,Shin2010}. While in STM and low energy electron diffraction (LEED) measurements almost no orientational preference of the islands was found, a theoretical analysis using Novaco-McTague theory indicated epitaxial growth with three main orientations of the islands~\cite{Shin2010}. This contradiction has been explained by an effective decrease in the C$_{60}$-graphite corrugation due to the vibrational and rotational motion of the C$_{60}$ molecules at room temperature. This decrease leads to an angular smearing in the observed orientations~\cite{Shin2010}.

We have studied the growth of C$_{60}$ on the HOPG surface using scanning tunneling microscopy (STM) at liquid nitrogen (LN$_2$) temperatures. We have found three possible orientations of the islands with respect to the substrate. We have obtained atomic resolution on HOPG and molecular resolution on C$_{60}$ within the same image. In this image we show that for a particular island there is an epitaxial relationship to the substrate. Two-dimensional Fourier transformation of high-resolution STM images supports this result.

\newpage

\section{\label{sec:exp}Experimental}
We used an approximately $ 3\times 13$~mm$^2$-sized piece of HOPG (grade $H$) with a mosaic spread of $3.5^\circ \pm 1.5^\circ$ (Optigraph GmbH). The HOPG sample was mounted on top of a Si(111) stripe that served as heater. The HOPG was cleaved in air and degassed at $T= 420-470$~K in ultra-high vacuum (UHV). The surface quality of the substrate was checked with low temperature STM prior to the molecule deposition, for details see supporting information. All STM experiments were performed in a low temperature STM (LT-STM, Omicron) at LN$_2$ temperatures in UHV. 
After checking the cleanliness of the HOPG surface, the substrate was transferred to an additional chamber of the system, the deposition chamber, and was thermalised in a room-temperature environment for $30$~min. 
The C$_{60}$ was deposited using a simple home-built evaporator. The evaporator consisted of a boron nitrate crucible that was heated by a stripe of tantalum foil. For evaporation the substrate was placed in front of the crucible at an angle of approximately $30^\circ$. We applied a constant current of $I=16.4$~A to the Ta-heater of the crucible and the amount of C$_{60}$ was controlled by varying the heating time between $1$~min $30$~s and $3$~min. For the first $1$~min $30$~s the amount of deposited C$_{60}$ remained negligible.
After the pressure had recovered, typically within $2-3$~min, the sample was directly transferred to the pre-cooled STM. For STM imaging the constant current mode was used for the distance controller. For some images, the tunneling current set point was varied during image acquisition as detailed below.
 
\section{\label{sec:res}Results and Discussion}
\begin{figure}\centering
\includegraphics[width=0.45\linewidth]{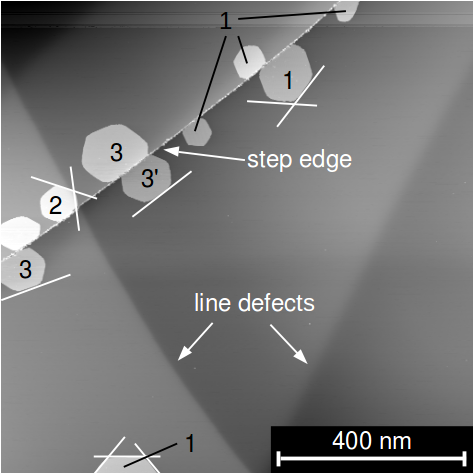}
\caption{Typical STM image of C$_{60}$ islands (evaporation time $1$~min $45$~s) obtained at a bias voltage of $U_{\mbox{\scriptsize B}} = 2$~V and a tunneling current of $I_{\mbox{\scriptsize T}} = 20$~pA. The HOPG step edge is decorated by islands while nearby line defects, presumably related to subsurface steps, are not. 
The islands have a hexagonal shape with rounded corners. Three distinct types of C$_{60}$ islands marked by the numbers 1 to 3 are found depending on the preferred orientations of their edges, which are indicated by white lines in the image. See text for details. \label{fig_1}}
\end{figure}

Fig.~\ref{fig_1} shows a typical LN$_2$-STM image of the HOPG surface after deposition of a small amount of C$_{60}$ molecules leading to a low coverage. Single molecular layer high C$_{60}$ islands cover roughly $5\%$ of the scanned surface. In most aspects the images agree well with previous STM studies~\cite{Szuba1999,Liu2006,Liu2008,Shin2010}.
The islands are preferentially arranged at step edges \cite{Burke2005,Burke2007,Kenny2000,Shin2010}. This preference for the growth at step edges is probably related to the high mobility of the C$_{60}$ on the graphite surface \cite{Shin2010}, and is in contrast with the absorption of C$_{60}$ on stronger interacting surfaces such as Pb/Si(111) \cite{Matetskiy2013} where C$_{60}$ typically nucleates at surface defects on terraces. Line defects attributed to subsurface HOPG defects are not decorated by C$_{60}$. The single molecular layer islands show hexagonally oriented
edges with rounded corners~\cite{Szuba1999}. 
The hexagonal orientation of the edges indicates the sixfold symmetry of C$_{60}$ in the direction perpendicular to the surface. This indication is confirmed by the molecularly resolved STM images discussed below. A hexagonal symmetry of C$_{60}$ islands is in agreement with previous studies of the growth of C$_{60}$ on graphite~\cite{Szuba1999,Liu2006,Liu2008,Shin2010}.
and also on CaF$_2$ below $300$~K~\cite{Koerner2011}.  

We have examined the preferred orientations of the hexagonal island edges to obtain information on the epitaxial relationship of the C$_{60}$ island to the substrate. The orientations marked with white lines in Fig.~\ref{fig_1} differ from island to island. For many islands, one of the preferred directions is nearly but not precisely parallel to the decorated HOPG step edge as observed in previous works~\cite{Shin2010}. The HOPG step in Fig.~\ref{fig_1} is slightly curved and the preferred hexagonal directions roughly follow its curved shape.
A detailed analysis of these orientations allows to divide the islands into three types. Type~1, defined by one edge of the island parallel to the $x$-axis of the image; type~2, defined by one edge rotated by approximately $30^\circ$ around the $x$-axis, i.e.\ one edge of the island parallel to the $y$-axis of the image; and type~3, defined by one edge that follows an orientation corresponding to neither of these two. In Fig.~\ref{fig_1} the edges of the type~3 islands form approximately an angle of $14^\circ$ with the ones of type~1 islands and an angle of $18^\circ$ with type~2 islands. The experimental error for these angles is quite large, since not always the island edges form a $120^\circ$ angle with respect to each other as it would be expected for perfect hexagons. Type~3' denotes a variant of type~3 obtained through a mirror symmetry of the type~3 islands with the mirror oriented along the $y$-axis of the image. 

\begin{figure}\centering
\includegraphics[width=0.9\textwidth]{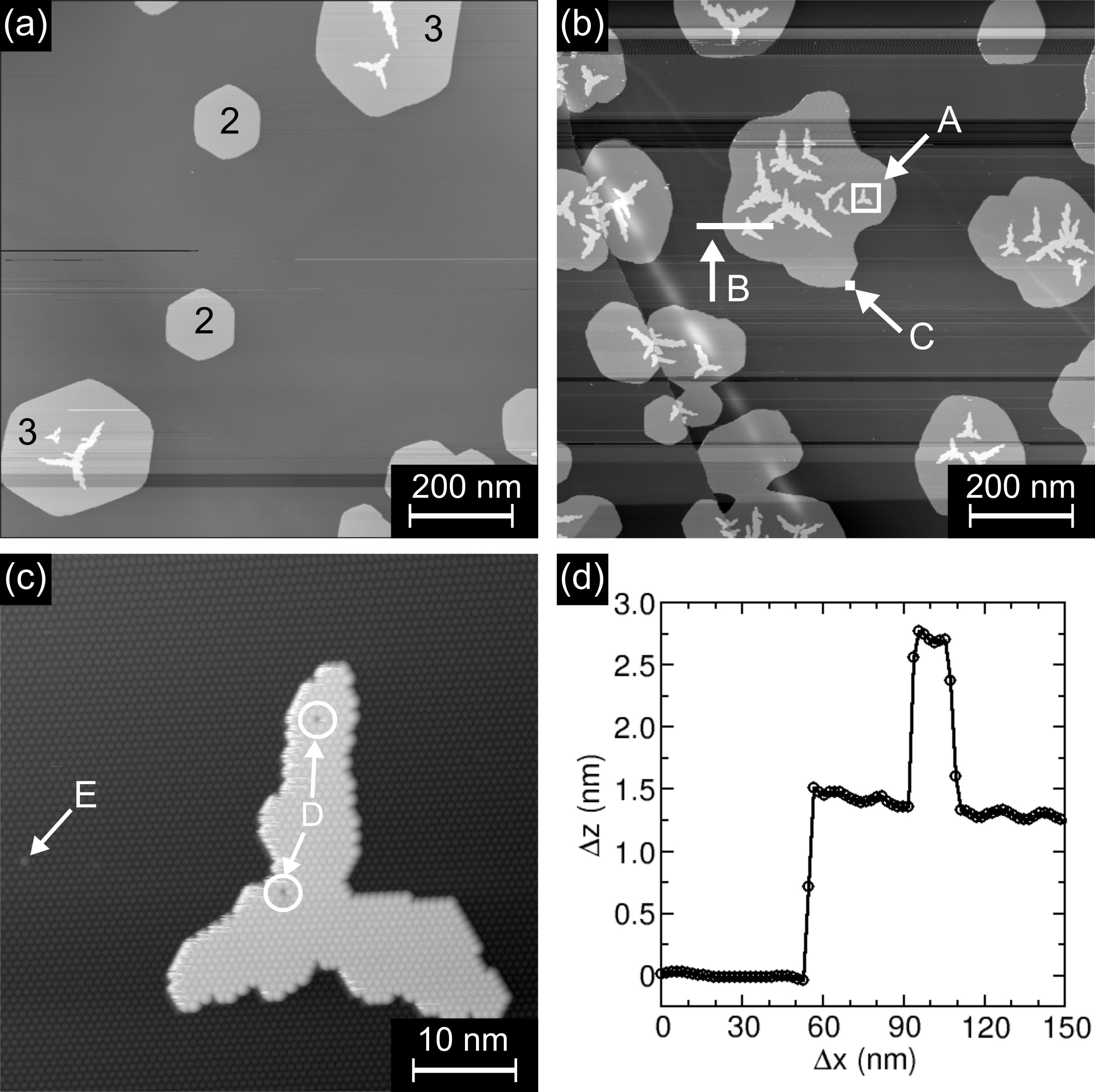}
\caption{STM image of C$_{60}$ islands on HOPG at a higher coverage compared to Fig.~\ref{fig_1}. a)~Onset of the second layer growth with a fractal-dentritic shape. Evaporation time $2$~min$ 15$~s, $U_{\mbox{\scriptsize B}} = 2$~V, $I_{\mbox{\scriptsize T}} = 20$~pA. b)~Larger islands often show flower-like shapes.
Within this area smaller images marked by A and C have been acquired as detailed below and in Fig.~\ref{fig_3}. Evaporation time $2$~min $30$~s, $U_{\mbox{\scriptsize B}} = 2$~V, $I_{\mbox{\scriptsize T}} = 20$~pA in lower part of the image, $I_{\mbox{\scriptsize T}} = 5$~nA in the upper third of the image. c)~Molecularly resolved image of the first and second layer of the C$_{60}$ island marked by A in~b). Several molecular-scale defects have been marked by D and E. $U_{\mbox{\scriptsize B}} = 2$~V, $I_{\mbox{\scriptsize T}} = 20$~pA. d)~Line profile measured along the line tagged B in~b).\label{fig_2}}
\end{figure}

Shin et al.~\cite{Shin2010} performed a detailed computational (Novaco-McTague) analysis of the orientation of the island's edges and found that $0^\circ$, $8^\circ$, and $29.9^\circ$ are stable orientations for the growth of C$_{60}$ on HOPG. However, their low-energy electron-diffraction and STM measurements only found a slight preference of the $30^\circ$-orientation. They explained this preference by the preference for the C$_{60}$ islands to nucleate and align at step edges~\cite{Shin2010} which are oriented at $30^\circ$ relative to the graphite lattice. Comparing these results to our data (see Fig.~\ref{fig_1} and Fig.~\ref{fig_2}~a)), we identify type~1 and type~2 islands with the structures observed at a $0^\circ$ and $29.9^\circ$. Type~3 and~3' islands respectively, although observed at an angle of $14^\circ$, could be tentatively identified with the $8^\circ$-structure within the experimental precision. 

Upon increasing coverage, both, the size and the number of the islands increase as shown in Fig.~\ref{fig_2}~a) and~b). For higher coverage islands additionally nucleate and grow on the terraces and not only at step edges. Probably the new nucleation centers correspond to residual impurities or defects on the surface \cite{Kenny2000}. In previous studies on graphene grown on 6H-SiC(0001), a substrate that also interacts weakly with C$_{60}$, the collective movement of small fullerene islands has been reported as a consequence of the interaction with the STM tip \cite{Svec2012}. The authors have observed that the islands that were not pinned by a defect showed a much faster mass transport. In our case larger islands for increasing coverage often show flower-like shapes instead of hexagonal ones, compare e.g. Fig. 2 a) and b). The flower-like shape is most likely the result of the coalescence of several hexagonal islands due to the large coverage (static coalescence \cite{Liu2006}). 

For increasing coverage also a second and higher layers grow on top of the single molecular layer high C$_{60}$ islands discussed above. 
In Fig. 2, we observe that the first layer, grown on HOPG, shows a compact growth while the second layer, grown on C$_{60}$, shows a fractal-dendritic growth in agreement with literature~\cite{Liu2006,Liu2008}. This is caused by a strong difference of the C$_{60}$ diffusion barrier between the HOPG substrate ($E_{\mbox{\scriptsize B}} = 13$~meV) and the C$_{60}$ island surface ($E_{\mbox{\scriptsize B}} = 168$~meV), suppressing step edge diffusion in the second layer, as suggested in~\cite{Liu2006,Liu2008,Gravil1996}. No indications of dewetting are observed. 
On insulator surfaces, in contrast, the C$_{60}$ molecules preferentially form a second layer and a peculiar island shape is caused by dewetting, such as in KBr(001)~\cite{Burke2007} and CaF$_2$ (111) surfaces~\cite{Koerner2011}. If the deposition time is increased further, the amount of layers and also the size of the layers both increase. Further measurements with variable substrate temperature during deposition should be performed to fully cover the dynamics of the growth.

\begin{figure}\centering
\includegraphics[width=0.45\linewidth]{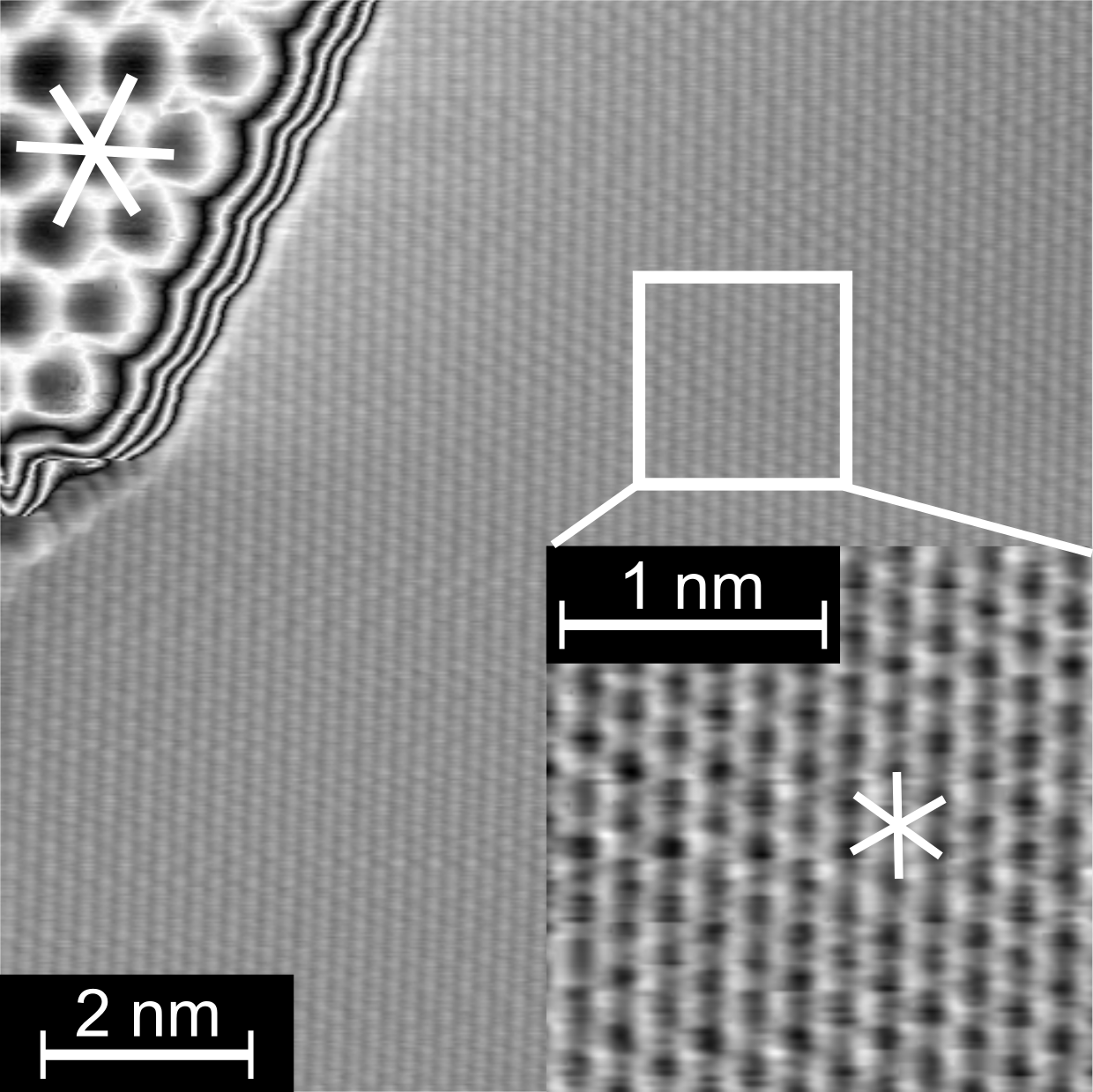}
\caption{High-resolution STM image at the area denoted as C in Fig.~\ref{fig_2}~b) comprising an edge of a type~1 island. The image shows molecular and atomic resolution on C$_{60}$ and HOPG, respectively within the same image. $U_{\mbox{\scriptsize B}} = 1.5$~V, $I_{\mbox{\scriptsize T}} = 20$~pA. A representation of the height values alternating five times between black and white has been chosen in order to show the high-resolution contrast, both, on the upper terrace of the C$_{60}$ island as well as on the HOPG substrate within the same image. Evaporation time $2$~min $30$~s. Inset: magnification of the $2\times 2$~nm$^2$ area marked with the white square. \label{fig_3}}
\end{figure}

After analysing the shape of the islands, we now concentrate on their internal structure. In Fig.~\ref{fig_2}~c) we show an enlarged  STM image  with molecular resolution of the region A marked with a square in Fig.~\ref{fig_2}~b). The C$_{60}$ island shows several defects labeled with arrows: D indicates two positions where a single C$_{60}$ molecule is missing in the top layer. A slightly brighter molecule is marked with E. The defect E could be caused by a defect in the HOPG surface, by a modified C$_{60}$ or by stress released due to the lattice mismatch between the C$_{60}$ and the HOPG surface.
Fig.~\ref{fig_2}~d) shows a line profile measured along the line tagged B in Fig.~\ref{fig_2}~b). The apparent height of the first layer is roughly $h_1=1.5$~nm. The apparent height measured with STM strongly varies in the literature between $h_1=1.05$~nm and $h_1=1.88$~nm~\cite{Szuba1999,Kenny2000,Liu2006}. In some cases, the large values are explained by an initial bilayer growth~\cite{Kenny2000}, while the low values of $h_1$ have been attributed to electronic effects~\cite{Liu2006}. 
Here, we assume a single molecule height. The deviation might be explained by imprecisions of the scanner calibration since we also measure an increased height for HOPG steps. HOPG steps appear a factor of about 1.5 higher in our images compared to literature. However, the calibration has been previously double-checked at liquid He temperatures on other surfaces such as Ag(111) and Nb(110) and appears to be precise \cite{Tomanic2012}. Therefore the deviation of the measured height for HOPG and C$_{60}$ layers on HOPG is most likely to be attributed to electronic effects of the surface and the measured voltage, since these materials are not simple metals. In addition, the influence of the electronic state of the tip has an influence on the STM measurements. The apparent height of the second layer is approximately $h_2=1.2$~nm. 
 
\begin{figure}\centering
\includegraphics[width=0.9\linewidth]{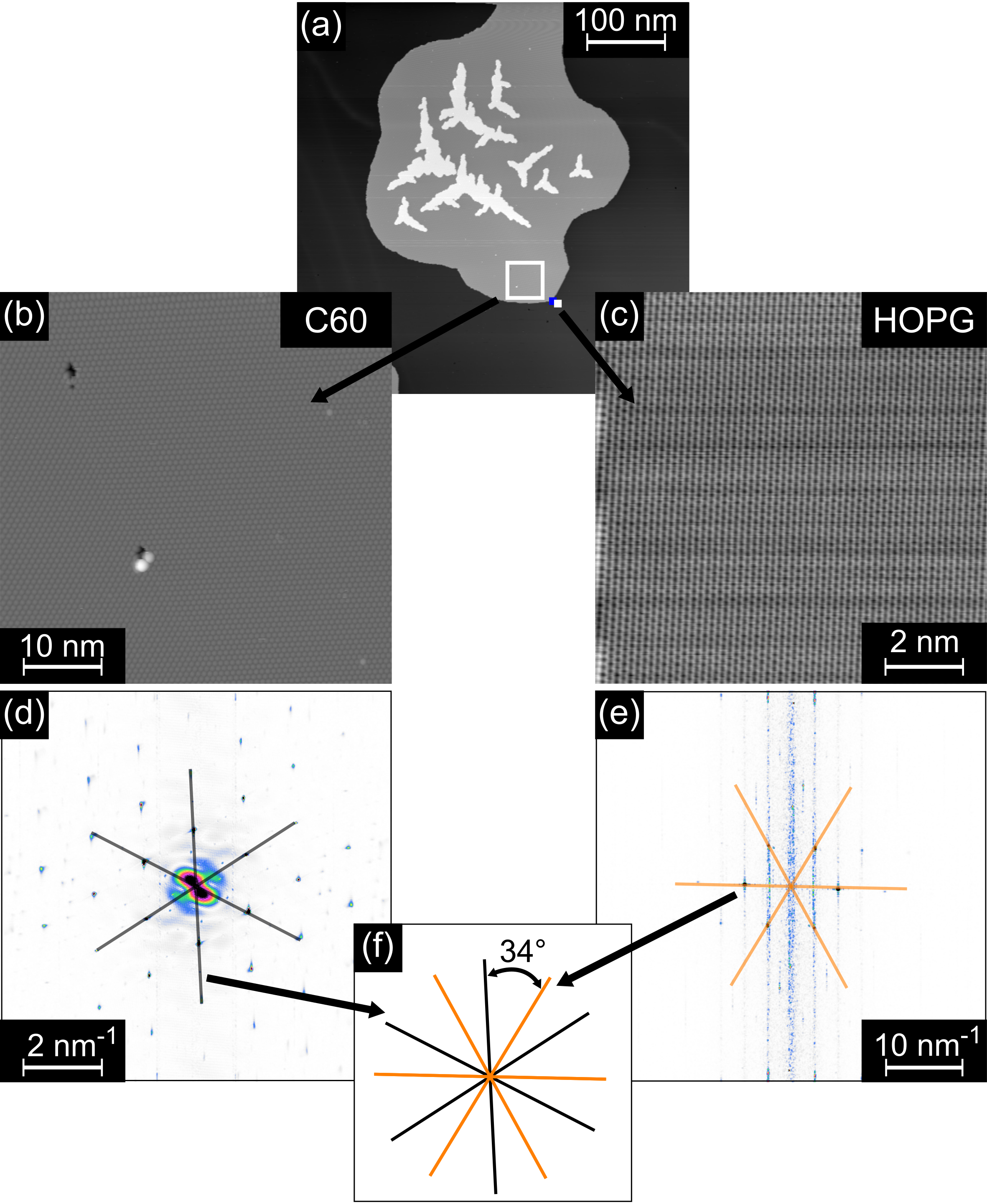}
\caption{2D-FFT analysis of molecularly and atomically resolved STM images. a)~STM image of the C$_{60}$ island in the center of Fig.~\ref{fig_2}~b). Scan size $500\times 500$~nm$^2$, $U_{\mbox{\scriptsize B}}=2$~V, $I_{\mbox{\scriptsize T}}=20$~pA. The region marked in blue corresponds to the area shown in Fig.~\ref{fig_3}, where we simultaneously obtained molecular an atomic resolution. b)~Area marked with an empty white square in~a) on the C$_{60}$ island. $50 \times 50$~nm$^2$, $U_{\mbox{\scriptsize B}}=2$~V, $I_{\mbox{\scriptsize T}}=20$~pA. c)~Area marked with a filled white square in~a) on the HOPG surface. $10 \times 10$~nm$^2$, $U_{\mbox{\scriptsize B}}=1.3$~V, $I_{\mbox{\scriptsize T}}=50$~pA. d)~2D-FFT data of image~b) obtained on the C$_{60}$ island, the main symmetry directions are marked in black. e)~2D-FFT data of image~c) obtained on HOPG, the main symmetry directions are marked in orange. f)~Comparison of the main symmetry directions of the C$_{60}$ island and the HOPG substrate. \label{fig_4}}
\end{figure}
 
To investigate the epitaxial relationship between the first layer of the C$_{60}$ island and the substrate, we have performed high-resolution imaging at the area denoted as C in Fig.~\ref{fig_2}~b) comprising an edge of a type~1 island. In this image shown in Fig.~\ref{fig_3}, we have obtained simultaneously atomic resolution on HOPG and molecular resolution on C$_{60}$. It is surprising that stable imaging conditions have been found in view of the strongly differing standard imaging conditions: atomic resolution on HOPG is usually obtained at a tunneling voltage of around $10-100$~mV while for high-resolution imaging of C$_{60}$ tunneling voltages of $1-2$~V are necessary~\cite{Liu2006,Shin2010}. This difference prevented such resolution in previous studies~\cite{Szuba1999,Liu2006,Liu2008,Shin2010}. The enhanced stability of our measurements could result from binding a C$_{60}$ molecule to the tip. Simultaneously imaging both materials reduces the influence of most artefacts typical in STM images~\cite{Burke2005}: imprecisions of the piezoelectric scanner calibration, piezoelectric hysteresis, creep, as well as thermal drift. By comparing the angles observed in the images we obtain an angle between the molecular layer and the substrate high-symmetry directions of approximately $30^\circ$ for this particular island. For the structure of type~1 islands, we have derived a ball model presented in the supporting information.

We have used two-dimensional fast Fourier transform (2D-FFT) of the STM images to obtain additional quantitative information about the epitaxial relationship between the C$_{60}$ islands and HOPG. Fig.~\ref{fig_4}~a) shows a small-scale image of the island in the middle of Fig.~\ref{fig_2}. This island has been also used for high-resolution imaging of Fig.~\ref{fig_3} (area marked in blue) as described above. In close vicinity to this area we have performed high-resolution imaging of an area fully covered by C$_{60}$, Fig.~\ref{fig_4}~b), and on the HOPG substrate, Fig.~\ref{fig_4}~c), with molecular and atomic resolution, respectively. The corresponding 2D-FFT data are shown in Fig.~\ref{fig_4}~d) and~e). The high symmetry directions of the C$_{60}$ lattice are marked in black and those of the HOPG surface in orange. Combining these two sets of directions in Fig.~\ref{fig_4}~f), we measure an angle of $34^\circ$ between them. The deviation of this angle with respect to the above-mentioned $30^\circ$ lies within the experimental error. The STM images have different sizes and different scanning speeds, they are differently affected by residual drift effects along slow and fast scanning axes possibly leading to small changes of the measured angles.

In literature the surface energy has been used as a measure of the molecule-substrate interaction in order to compare the formation of C$_{60}$ islands on wide-band gap insulators~\cite{Rahe2012}. Due to the important molecular mobility, kinetic barriers can be overcome at the temperature at which the islands are formed. This also applies to HOPG substrates, where the weak van-der-Waals interaction favors island mobility \cite{Liu2008,Svec2012}. The idea has been put forward that low surface energy materials show dewetting in contrast to high surface energy materials~\cite{Rahe2012}. This idea relates to the well-known general considerations that growth modes depend on the surface and interface energies~\cite{Venables2000}.
C$_{60}$ shows dewetting for KBr~\cite{Burke2005}, the insulator surface with the lowest surface energy available in that study ($141$~mJ/m$^2$~\cite{Rahe2012}), while it shows wetting on CaCO$_3$(10$\overline{1}$4) with a much higher surface energy ($590$~mJ/m$^2$\cite{Rahe2012}). On graphite, even though graphite has an even lower surface energy compared to KBr ($54.8$~mJ/m$^2$~\cite{Wang2009}), C$_{60}$ shows wetting comparable to the observations on CaCO$_3$(10$\overline{1}$4).
This observation suggests that the surface energy alone cannot fully
describe the interaction of molecules with a surface. The detailed
nature of the interaction, e.g. whether the substrate is a metal or an insulator, or potentially the interaction energy of the molecule with the surface should be considered rather than the surface energy of the substrate alone. It appears that the surface energy of the substrate is only a well-chosen parameter if the same type of materials are considered, e.g.\ for comparison of different ionic crystalline surfaces. 

\section{\label{sec:conc}Conclusion}
We have investigated the growth of C$_{60}$ on HOPG using STM at liquid nitrogen temperatures. For low coverage C$_{60}$ single molecular layer islands grow in the form of hexagonally oriented edges with rounded corners decorating HOPG step edges. Upon increasing the coverage flower-like shape islands also grow on terraces, and multilayers appear showing a fractal-dendritic shape. In contrast to previous results, in our study we find three different orientations of the islands' edges as predicted by theory. The remarkable stability of the islands have permitted further investigations. For one particular island, we have studied the epitaxial relationship with the substrate by analyzing a high-resolution image across an island edge where we have obtained molecular and atomic resolution within the same image.
A 2D-FFT analysis of STM images on this region confirms that the symmetry directions of the C$_{60}$ islands and the HOPG surface form an angle of approximately $30^\circ$.

\section{\label{sec:ackn}Acknowledgment}
This work was supported by the Alexander von Humboldt-foundation, and the ERC Starting Grant NANOCONTACTS. 

\section*{References}

\newpage

\section{\label{sec:si}Supporting Information}

\subsection{\label{sec:si:clean}Clean HOPG surface}
\begin{figure}[h]\centering
\includegraphics[width=0.45\linewidth]{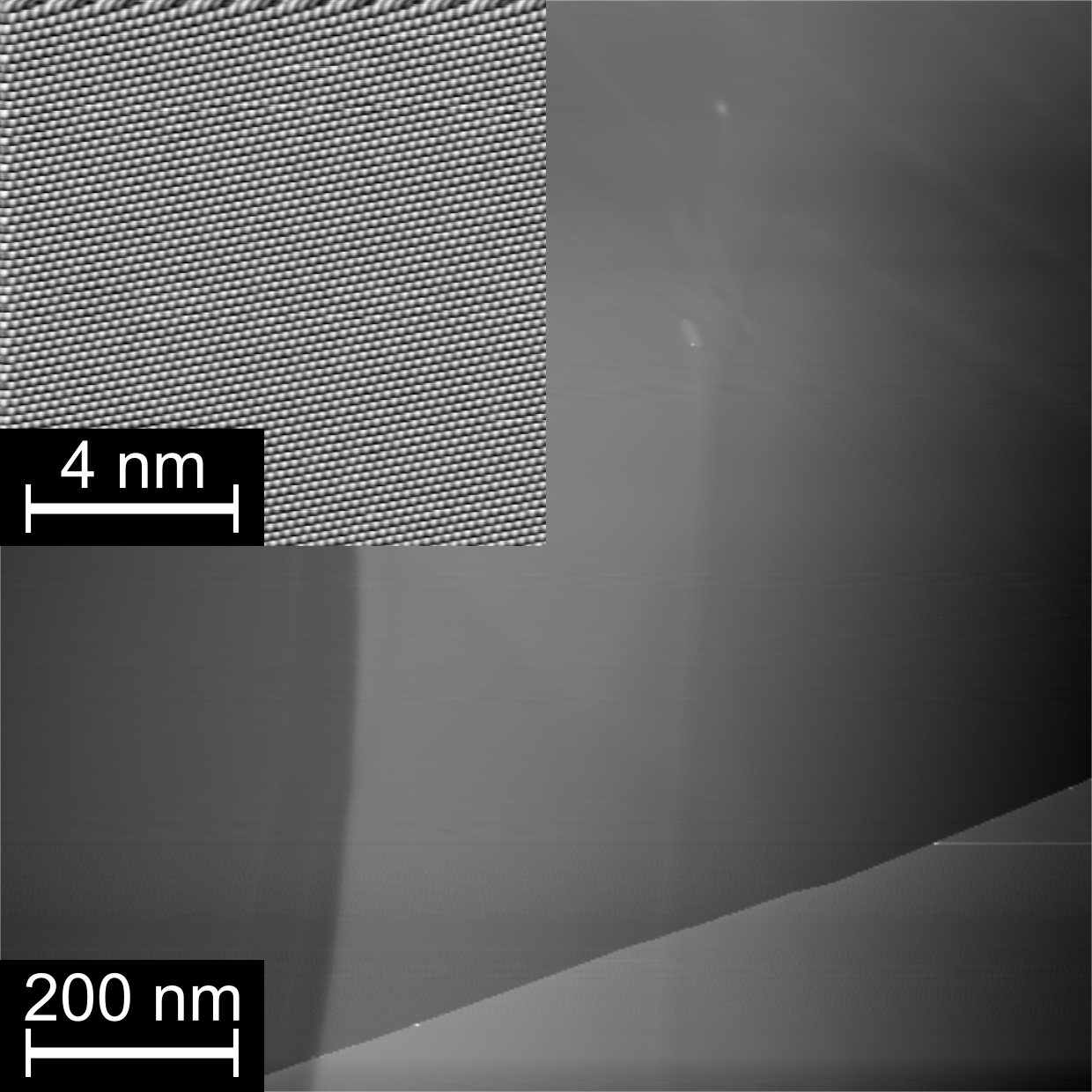}
\caption{Clean HOPG surface after cleavage. Scan size $1 \times 1 \mu$m$^2$. Typical terasses show a width of several hundreds of nanometers. $U_{\mbox{\scriptsize B}}= 2$~V, $I_{\mbox{\scriptsize T}}= 20-250$~pA. The inset depicts a typical image with atomic resolution for HOPG. $ U_{\mbox{\scriptsize B}}= 0.2$~V, $I_{\mbox{\scriptsize T}} = 250$~pA. \label{fig_5}}
\end{figure}

\subsection{\label{sec:si:model}Ball-model for C$_{60}$ on HOPG}
In order to investigate the epitaxial relationship in more detail we have extended manually the molecular arrangement found in Fig.~\ref{fig_3} until it overlapped to the also extended atomic positions of the HOPG surface taken from the same image. Although the image is slightly distorted in this region, the angles of the high-symmetry lines give a sensitive measure of the atomic/molecular arrangement and allow to precisely continue the pattern. With this comparison, we have generated a model of the epitaxial relationship. We obtain a commensurate superstructure within the experimental precision. As it is typical for molecular molecular epitaxy, the superstructure unit cell is large compared to the substrate unit cell~\cite{Hooks2001}.

\begin{figure}
\centering
\includegraphics[width=0.7\linewidth]{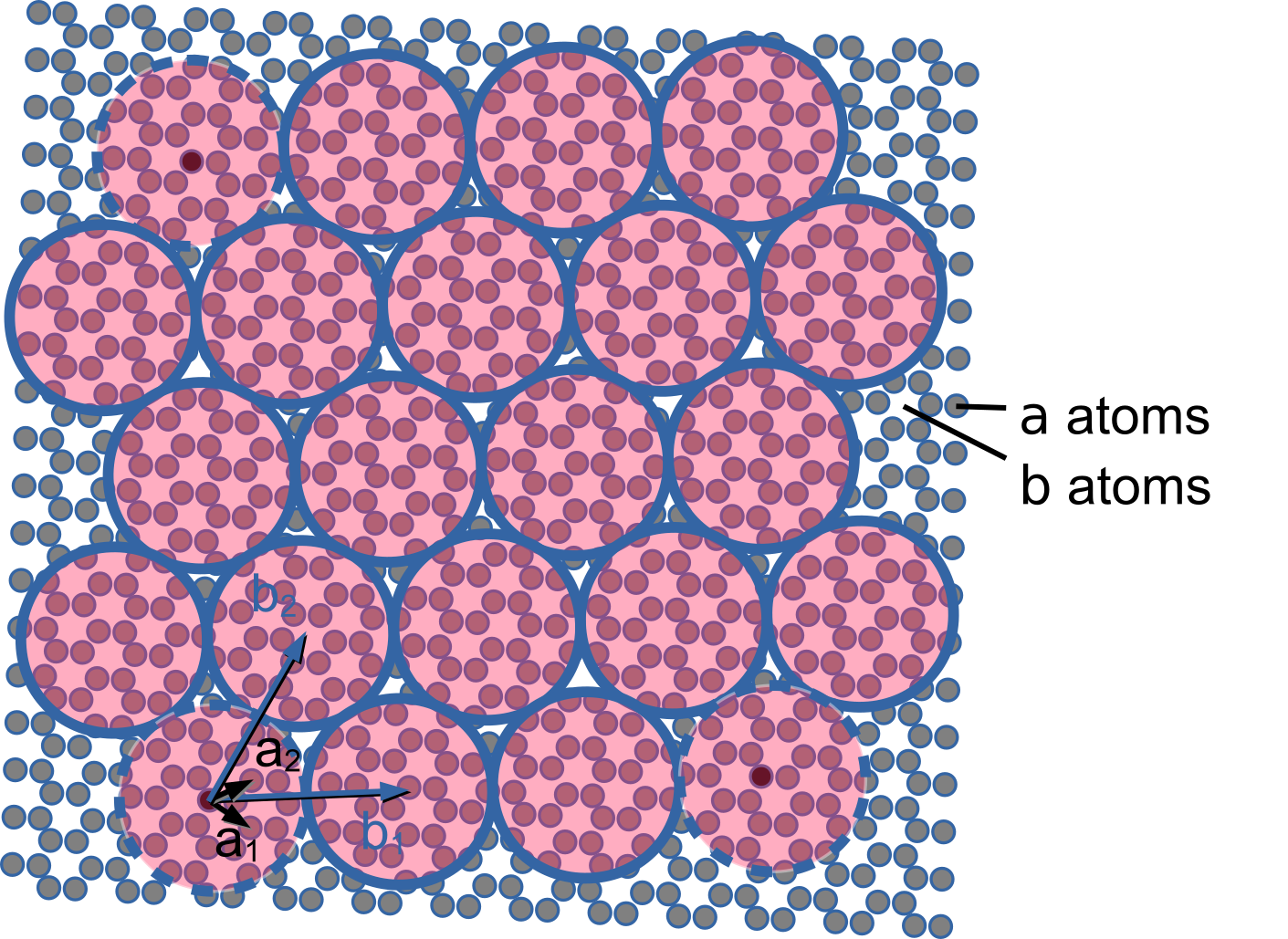}
\caption{
Ball-model for C$_{60}$ on HOPG showing the structure of a type~1 island
derived from the STM image of Fig.~\ref{fig_3}.\label{fig_6}}
\end{figure}

Taking the graphite-unit cell as a basis and defining a superstructure by
\begin{equation*} 
\left(
\begin{array}{c}
\vec{b}_1\\
\vec{b}_2\\
\end{array}
\right)
=
\left(
\begin{array}{cc}
c_{11} & c_{21}\\
c_{12} & c_{22}\\
\end{array}
\right)
\cdot
\left(
\begin{array}{c}
\vec{a}_1\\
\vec{a}_2\\
\end{array}
\right)
\end{equation*}
with the definition of the lattice vectors as defined in Fig.~\ref{fig_6}, we determine an approximate superstructure of
\begin{equation*} 
\left(
\begin{array}{c}
\vec{b}_1\\
\vec{b}_2\\
\end{array}
\right)
=
\left(
\begin{array}{cc}
2 & 2.67\\
-2.5 & 4.58\\
\end{array}
\right)
\cdot
\left(
\begin{array}{c}
\vec{a}_1\\
\vec{a}_2\\
\end{array}
\right)
\end{equation*}
For a precise commensurate structure, the C$_{60}$ island is slightly compressed and/or rotated with respect to the substrate. This has not been taken into account in the model shown in Fig.~\ref{fig_6}.

\end{document}